\documentstyle[11pt]{article}
\begin{document}

\def\beq{\begin{equation}}
\def\eeq{\end{equation}}
\def\ul{\underline}
\def\ni{\noindent}
\def\nn{\nonumber}
\def\wt{\widetilde}
\def\wh{\widehat}
\def\Tr{\mbox{Tr}\ }
\def\tr{\,\mbox{tr}\,}                  
\def\Tr{\,\mbox{Tr}\,}                  
\def\Res{\,\mbox{Res}\,}                
\renewcommand{\Re}{\,\mbox{Re}\,}       
\renewcommand{\Im}{\,\mbox{Im}\,}       
\def\lap{\Delta}                        

\def\al{\alpha}
\def\be{\beta}
\def\ga{\gamma}
\def\de{\delta}
\def\ep{\varepsilon}
\def\ze{\zeta}
\def\io{\iota}
\def\ka{\kappa}
\def\la{\lambda}
\def\na{\nabla}
\def\ro{\varrho}
\def\si{\sigma}
\def\om{\omega}
\def\ph{\varphi}
\def\th{\theta}
\def\te{\vartheta}
\def\up{\upsilon}
\def\Ga{\Gamma}
\def\De{\Delta}
\def\La{\Lambda}
\def\Si{\Sigma}
\def\Om{\Omega}
\def\Te{\Theta}
\def\Th{\Theta}
\def\Up{\Upsilon}

\hfill 8 of Augest, 1996


\vspace*{3mm}

\begin{center}

{\Large \bf On the conformal transformation and duality in gravity}

\vspace{15mm}


{\sc Ilya L. Shapiro} \footnote{On leave from Tomsk Pedagogical 
University, 634041, Tomsk, Russia \\
$\,\,\,\,\,$ E-mail: shapiro@fisica.ufjf.br}\\

\vskip 5mm

Departamento de Fisica Teorica, Universidad de Zaragoza, 50009, Spain
\\
and
\\
Departamento de Fisica -- ICE,
Universidade Federal de Juis de Fora -- MG, Brazil
\vskip 10mm

\vspace{15mm}
\end{center}

\noindent
{\large \sl Abstract.} $\;\;\;$
The theory described by the sum of the Einstein-Hilbert action 
and the action of conformal scalar field possesses the duality 
symmetry which includes some special conformal transformation 
of the metric, and also inversion of scalar field and of the 
gravitational constant. In the present paper the conformal duality 
is generalized for an arbitrary space-time dimension $n \neq 2$ and 
for the general sigma-model type conformal scalar theory. We also 
consider to apply the conformal duality for the investigation of 
quantum gravity in the strong curvature regime. The trace of the 
first coefficient of the Schwinger-DeWitt expansion is derived 
and it's dependence on the gauge fixing condition is considered. 
After that we discuss the way to extract the gauge-fixing 
independent result and also it's possible physical applications.

\vspace{20mm}

\noindent{\large \bf Introduction}
\vskip 1mm

It is well known that the study in the field of quantum gravity
meets serious difficulties. On classical level General Relativity
is in a good accordance with all the known tests, but
quantum theory based on GR is nonrenormalizable
\cite{hove}. Indeed there are
a number of approaches to investigate some aspects of
quantum gravitational phenomenons. One can mention the idea of inducing GR
from the quantum effects of matter fields or from the theory of string
(see, for example, \cite{adl} and \cite{gsw}). Recently there were a very
interesting attempts to extract the low-energy predictions from the
GR-based quantum theory of gravity \cite{don,tswo}.
However the principal problem with the lack of quantum gravitational 
experiments doesn't 
enable one to choose among the numerous theories and therefore the
models for quantum gravity are subjects of wide arbitrariness.
This concerns, for instance, the choice of the model for the
description of low-energy quantum gravitational phenomena, where the
string-inspired action contains, along with the metric, massless
dilaton field.
In such a situation a lot of attention has been attracted by the
models which have more symmetries than GR. In particular, one can
mention the supergravity theories, and also the theories with local
conformal invariance. The conformal transformation and conformal
symmetry in gravity are especially
interesting in many respects \cite{des,bek} because of their relation with
the cosmological applications \cite{bar}
and with the physically important issues
of renormalization group \cite{cthe},
conformal anomaly and anomaly induced gravitational effective
action \cite{20let,rei,bos,anmo,cosh,osborn}.

Another kind of problem concerns the methods of
calculations in quantum gravity. The standard perturbative techniques
give the effective action in a form of
a power series in the curvature tensor and it's
derivatives (see \cite{DTGF} and \cite{book} for the review).
At the same time one of the most important
applications of quantum gravity should be the regions of extremely
strong gravitational field in the vicinity of cosmological and
black hole singularities. One way to study this subject is to develop
essentially nonperturbative methods in quantum gravity like dynamical
triangulations which have been recently applied to the four
dimensional gravity \cite{amb}. Another one is to explore a specific
models for quantum gravity which have a symmetry linking
the regimes of strong and weak gravitational field. In this paper
we consider an example of such a symmetry -- conformal duality, which
has been originally discovered by Bekenstein \cite{bek} as a property
of dynamical equations for the conformal scalar field coupled to gravity.

Our purpose is to investigate the conformal duality 
on quantum level, and hence we face the problem of nonrenormalizability
of quantum gravity. The unitary second derivative models of quantum
gravity are nonrenormalizable already at one loop if they are considered
together with the matter fields \cite{hove,dene}. In fact the one-loop
approximation has a special significance. If considering
the asymptotically free theory\footnote{One can suppose that the
renormalization group for gravity can be formulated in the
nonperturbative way, and so can be applied even for the
nonrenormalizable theories \cite{wei}}, then the one-loop approximation
provides the most of the information about the effective action
in the strong gravitational field regime (which is natural $UV$ limit
for the nontrivial background metric \cite{book}).
One of the possible ways to avoid the problem of divergences
is to consider the space-time dimension $n$ different from four.
It is well known that in some dimensions the divergences can be kept
under control, at least in some approximation. For instance,
in any odd dimension the divergences
are lacking at the one-loop level \cite{dew}. That is why in
this work we start with generalizing the conformal duality of
\cite{bek} for the most general second derivative metric-dilaton model
in almost arbitrary space-time dimension $n \neq 2$. For this purpose
we use the action instead of dynamical equations \cite{conf} that is
easier and facilitates the consequent quantum consideration, which
is based on the lower orders of the Schwinger-DeWitt expansion of the 
effective action.

The paper is organized as follows.
In the next section we write the actions and symmetry transformations
for the conformal metric-scalar theory and for the theory with
conformal duality in arbitrary dimension $n$. Section 3 is devoted to
the calculation of the first term in the Schwinger-DeWitt expansion
of the one-loop effective action. In section 4 we consider the 
gauge-fixing
dependence of the one-loop effective action 
and also construct the gauge-independent combination of it's components.
In the last section some conclusions and also discussion of the next
reasonable steps in the study of conformal duality are presented.
Some information which is aimed to support the main text, is placed in
the Appendixes.
The possibility of the soft breaking of conformal
symmetry is discussed in Appendix A, and the equations of motion for
the dual model are written down in Appendix B, where we also solve the
equation for the time-dependent dilaton. Both conformal and
dual actions are singular in the limit $n \rightarrow 2$, which is
explored in some details in Appendix C. Appendix D contains bulky
expansion of the dual action which has been used in the one-loop
calculations.

\vspace{5mm}

\noindent{\large \bf 2. General theory with conformal symmetry and
duality.}
\vskip 1mm

We consider the second-derivative metric-dilaton
model with the action of sigma-model type in $n$-dimensional 
space-time.
\beq
S[g_{\mu\nu}, \phi] =
\int d^nx \sqrt{-g}\; 
\left\{ \, A(\phi)\,g^{\mu\nu}\partial_{\mu}\phi\,
\partial_{\nu}\phi + B(\phi)\,R + C(\phi) \, \right\}     \label{0.1}
\eeq
This theory possesses general covariance and, for some special choice
of the functions $A,B,C$ (otherwise arbitrary), an extra conformal
symmetry. As a first step we describe the local conformal
transformation in the theory (\ref{0.1}) without any restrictions for
these functions.
For this purpose we need the relations between geometric quantities
related with the metric $g_{\mu\nu}$ and with the transformed metric
${\bar g}_{\mu\nu}=g_{\mu\nu}\;e^{2\si}$, where $\si = \si (x)$.
\beq
\sqrt{-{\bar g}} = \sqrt{-g}\;e^{n\si}
,\;\;\;\;\;\;\;\;\;\;\;\;\;\;\;\;
{\bar R} = R({\bar g}_{\mu\nu}) = e^{-2\si}\left[R - 2(n-1)(\Box\si)
-(n-2)(n-1)(\na\si)^2 \right]                      \label{n1}
\eeq
Substituting (\ref{n1}) into (\ref{0.1}) we find
$$
S[{\bar g}, \phi] =
\int d^nx \sqrt{-g}
\left\{	e^{(n-2)\si} \left[
A(\na \phi)^2 + (n-1)(n-2) B(\na\si)^2 +
\right.\right.
$$
\beq
\left.\left.
+ 2(n-1)B_1(\na^\mu\phi)(\na_\mu\si) +
R B\right] + e^{n\si}C \right\}                           \label{n2}
\eeq
where we have used the notations of \cite{conf}:
$ X_i = {d^i X} / {d \phi^i},\;\;\;\;
(\na X)^2 = g^{\mu\nu}\;\partial_\mu\;X\;\partial_\nu\;X $.

If one performs an arbitrary reparametrization of the scalar field
$\phi = \phi(\psi)$, then the action becomes
$S[{\bar g}_{\mu\nu}, \psi]$. The condition of symmetry for the
action (\ref{0.1})
$$
S[{\bar g}_{\mu\nu}, \psi] = S[g_{\mu\nu}, \phi]
$$
can be rewritten as the equations for the functions $A(\phi), B(\phi),
C(\phi)$ and for $\sigma(\phi)$.
$$
B(\phi) = e^{(n-2)\si}\;B\left(\psi(\phi)\right)
$$
$$
A(\phi) = A\left(\psi(\phi)\right) \left( \frac{d\psi}{d\phi} \right)^2 +
(n-1)(n-2) B\left(\psi(\phi)\right) \left( \frac{d\si}{d\phi} \right)^2 +
2 (n-1) \left( \frac{d\psi}{d\phi} \right)
\left( \frac{d\si}{d\phi} \right)
\left( \frac{dB(\psi)}{d\psi} \right)_{\psi = \psi(\phi)}
$$
\beq
C(\phi) = e^{n\si}\;C\left(\psi(\phi)\right)              \label{n3}
\eeq
The solution for these equations has the form
\beq
e^{(n-2)\si(\phi)} = \frac{B(\phi)}{B\left(\psi(\phi)\right)}    
\label{n4}
\eeq
\beq
A(\phi) = \frac{n-1}{n-2}\;\frac{{B_1}^2(\phi)}{B(\phi)}
\,,\;\;\;\;\;\;\;\;\;\;\;\;\;\;\;\;\;\;\;\;\;\;\;\;\;\;\;
C(\phi) = \la \;B^{\frac{n}{n-2}}(\phi) \, ,
\;\;\;\;\;\;\;\;\;\;\;\;\;\;\;
\la = const                                               \label{n6}
\eeq
Indeed the equations (\ref{n4}), (\ref{n6}) have different sense.
(\ref{n4}) shows the relation between arbitrary reparametrization of
scalar field and the corresponding conformal transformation for the metric.
Therefore the symmetry we have found is nothing but the direct
generalization
of the ordinary one-parameter conformal symmetry. The equations
(\ref{n6}) give the constraints on $A(\phi),\,C(\phi)$ which are caused
by conformal symmetry. In the Appendix A it is shown that the dynamical
equations for the theory with
$A(\phi)$ satisfying (\ref{n6}) and $C(\phi)$ not, do not have
any solutions.
It is useful to introduce the special notation for the action (\ref{0.1})
which satisfies the conformal constraints (\ref{n6}).
Following \cite{conf} we denote the action of such theory as
$S_{B(\phi),\la}$. One can easily see that one of the particular cases of
$S_{B(\phi),\la}$ is the Einstein gravity with
\beq
B = {\ga}\ka^{-2}                                \label{n7}
\eeq
Here $[\ka^2]$ is universal dimensional constant, and $\ga$ is some
dimensionless one. We split the Newton constant into two parts
for convenience. $C$ term here becomes cosmological constant.

Another particular case of $S_{B,\la}$ is the conventional conformal
scalar field.
One can start with arbitrary $S_{B(\phi),\la}$ and put
\beq
B(\phi) = \frac{n-2}{2(n-1)}\psi^2                \label{n8}
\eeq
where $\psi$ is some other scalar. Then, solving (\ref{n8}) in the form
$\phi = \phi(\psi)$ we find the reparametrization which links an
arbitrary $S_{B(\phi),\la},\;\; B(\phi) \neq const$ with particular
model of conventional conformal scalar.
\footnote{Of course some sigh restrictions
for the possible form of $B(\phi)$ appears
(see second reference in \cite{des}), and for some cases one needs
to change the sign in (\ref{n8}), otherwise the correspondence with
(\ref{n7}) can not be achieved.}
The only one sort of $S_{B(\phi),\la}$
which can not be reparametrized in this way is Einstein-Hilbert action
(\ref{n7}).

In fact GR is also equivalent to conformal scalar field \cite{des}.
To see this we substitute the conditions
(\ref{n6}) into the transformed expression (\ref{n2}). It turns
out that the last also satisfies (\ref{n6}) so that
\beq
S_{B(\phi),\la}[{\bar g}_{\mu\nu}, \phi] =
S_{K(\phi),\la}[g_{\mu\nu}, \phi]               \label{n9}
\eeq
for any given function $K(\phi)$ (with only sign restrictions), 
if only
\beq
{\bar g}_{\mu\nu} =
\left[\frac{K(\phi)}{B(\phi)}\right]^{\frac{2}{n-2}}\;g_{\mu\nu} 
\label{n10}
\eeq
In particular one can choose $K = const$ and demonstrate the conformal
equivalence between (\ref{n7}) and other models $S_{B(\phi),\la}$
with nonconstant $B$.

The equality (\ref{n9}) can be used as a basis for another interesting
symmetry, which holds for some of the nonconformal versions of 
(\ref{0.1}). Let us consider the sum
\beq
S_{B(\phi),\la}[{\bar g}_{\mu\nu}, \phi] +
S_{L(\phi),\tau}[{\bar g}_{\mu\nu}, \phi]                 \label{n11}
\eeq
where $B(\phi), L(\phi)$ are some functions and $\la, \tau$ are some
(arbitrary) constants. Perform  the special conformal transformation
(\ref{n10}) and find that (\ref{n11}) becomes
\beq
S_{K(\phi),\la}[g_{\mu\nu}, \phi] +
S_{M(\phi),\tau}[g_{\mu\nu}, \phi],
\;\;\;\;\;\;\;\;\;\;\;\;\;\;\;\;\;\;\;\;\;\;\;\;\;\;\;\;
M(\phi) = \frac{L(\phi)K(\phi)}{B(\phi)}                        
\label{n12}
\eeq
Especially interesting particular case of the above symmetry takes place
when one of the components in (\ref{n11}) is
GR $\,\,L = {\ga}\ka^{-2} = const$. For the sake of
simplicity we choose the second function to be
$B(\phi) = {\phi}\ka^{-2}$, therefore scalar is dimensionless.
Then the symmetry transformation reads
\beq
\left\{ S_{\frac{\phi}{\ka^2},\la} +
S_{\frac{\ga}{\ka^2},\tau}\right\}_{g_{\mu\nu}}
=  
\left\{ S_{\frac{1}{\phi\ka^2},\tau} +
S_{\frac{1}{\ga\ka^2},\la}\right\}_{{\bar g}_{\mu\nu}}       \label{n13}
\eeq
and hence for this specific case we face a conformal duality which
exchanges
$$
\phi \leftrightarrow {\phi}^{-1}
,\;\;\;\;\;\;\;\;\;\;\;\;\;\;\;\;\;\;
\ga \leftrightarrow {\ga}^{-1}
,\;\;\;\;\;\;\;\;\;\;\;\;\;\;\;\;\;\;
\tau \leftrightarrow \la
$$
\beq
{\bar g}_{\mu\nu} \leftrightarrow g_{\mu\nu},
\;\;\;\;\;\;\;\;\;\;\;\;\;\;\;\;\;\;{\bar g}_{\mu\nu} =
g_{\mu\nu}\;\left( {\phi\ga}\right)^{- \frac{2}{n-2}}  \label{n14}
\eeq
The equations of motion for the theory (\ref{n13}) have some 
interesting features, which are considered in Appendix B.
One can see that the conformal solution (\ref{n6}) and (consequently)
the action of the theory with conformal duality
become singular when $n$ tends to $2$. Some discussion of this limit is
given in Appendix C.

The dual symmetry links different metrics, different values of
scalar field and different values of coupling constants. Let us, for
instance, choose weakly changing $\phi << 1$ and also $\ga << 1$. Then
the curvatures of two metrics ${\bar g}_{\mu\nu}$ and
$g_{\mu\nu}$ are linked by relation (\ref{n1}).
If one supposes that the last terms in the brackets in (\ref{n1})
are of the same order or smaller than the scalar
curvature, we arrive at the transformation which links the
regimes of strong and weak gravitational field. The metric
$g_{\mu\nu}$ can be almost flat and it's curvature very small while
the curvature of the metric ${\bar g}_{\mu\nu}$ can have a big value.
Simultaneously one meets a big value of the scalar $(1 / \phi)$ and a
of the dimensionless constant $(1 / \ga)$.

This physical interpretation of conformal duality looks different 
from the one which is common for other types of dual symmetries
(see, for example, the review \cite{dual}). Contrary to the string
dualities, the conformal duality does not link the regimes of
weak and strong coupling, because the coupling constant $\ka$
which is the parameter of the loop expansion in the path
integral, does not change under the conformal duality
transformation.

However the conformal duality can be important in the study of quantum
gravity effects in the regime of strong gravitational field in a given
order of the loop expansion. To illustrate this, consider the one-loop
effective action in the theory of quantum gravity with classical action
$S[g_{\mu\nu}, \phi]$. The effective action $\Ga$ can be
derived on the basis of the background field method and Schwinger-DeWitt
expansion. The last gives the local representation of $\Ga$ as an 
infinite power series in the proper time parameter $s$. Since this
parameter is dimensional, the local $\Tr a_k(x,x)$ coefficients have
dimensions of $(curvature)^k$, so such an expansion can be efficient for
the weak gravitational field only. Below we derive the first coefficient
$\Tr a_1(x,x)$ coefficient for the theory with conformal duality
and investigate it's gauge dependence.
\vspace{5mm}

\noindent{\large \bf 3. One-loop calculation}
\vskip 1mm

The purpose of this section is to formulate the one-loop effective
action for the theory with conformal duality (\ref{n13}) and to
evaluate the first nontrivial term in the Schwinger-DeWitt expansion
for the effective action in an arbitrary dimension.
Since all the theories with conformal duality (\ref{n12}) differ from the
most simple one (\ref{n13}) by the reparametrization of the scalar field
only, we shall perform the calculations for this simple case and thus
start with the theory
\beq
S = \frac{1}{\ka^2}\;\int d^nx \sqrt{-g}\; \left\{ \,\frac{n-1}{n-2}
\;\frac{1}{\phi}\;g^{\mu\nu}\partial_{\mu}\phi\;
\partial_{\nu}\phi + (\phi + \ga)R + V(\phi) \, \right\}     \label{2.1}
\eeq
where $V(\phi) = \la \phi^{\frac{n}{n-2}} + \tau \ga^{\frac{n}{n-2}}$.

For the sake of quantum calculations we apply the background field method
(see \cite{book} for the introduction).
The features of the metric-dilaton theory require the 
special background gauge, which has been originally
introduced in the similar two-dimensional theory \cite{odsh}
and recently applied for the calculation of the one-loop divergences
in general four-dimensional metric-scalar
theory (\ref{0.1}) in \cite{spec}.
The starting point of the calculations is the usual splitting of the fields
into background $W^n = \left( g_{\mu\nu}, \phi \right)$ and quantum
$w^n = \left( {\bar h}_{\mu\nu}, h, \varphi \right)$ ones as
\beq
\phi \rightarrow \phi' = \varphi + \ka\,\phi,              \;\;\;\;\;\;\;\;
\;   g_{\mu\nu} \rightarrow g'_{\mu\nu} + \ka\,h_{\mu\nu}, \;\;\;\;\;\;\;\;
h_{\mu\nu} = {\bar h}_{\mu\nu}+\frac{1}{n}\; g_{\mu\nu}h,\;\;\;\;\;\;\;\;
h=h_{\mu}^{\mu} \label{2.2}
\eeq
where the trace and the traceless parts of the quantum metric have been
separated for convenience. The details of expansion of the action
(\ref{2.1}) into series one can find in the Appendix D. The one-loop
effective action is given by the standard general expression
\beq
\Gamma^{1-loop}={i \over 2}\;\Tr\ln{\hat{H}}-i\;\Tr\ln {\hat{H}_{ghost}}
+ {i \over 2}\;\Tr\ln{Y^{\mu\nu}}                       \label{2.3}
\eeq
where $\hat{H}$ is the Hermitean bilinear form of the action 
$S + S_{gf}$ with added gauge fixing term
\beq
S_{gf} = \int d^n x \sqrt{-g}\;\chi_{\mu}\;Y^{\mu\nu}\;\chi_{\nu},
\label{2.4}
\eeq
which contains a weight function $Y^{\mu\nu}$.
$\hat{H}_{ghost}$ is the bilinear form of the action of the gauge ghosts.
The general form of the
gauge fixing condition and weight function are\footnote{We consider the
linear covariant background gauges only.}
\beq
\chi_{\mu} = \nabla_{\la} {\bar h}_{\mu}^{\,\la} +
\beta \,\nabla_{\mu}h + \rho \,\nabla_{\mu} \varphi,
\;\;\;\;\;\;\;\;\;\;\;\;\;\;\;\;\;\;\;\;\;\;\;
Y^{\mu\nu} = - \al \, g^{\mu\nu}
 \label{2.5}
\eeq
where the gauge fixing parameters
$\alpha, \beta, \rho$ are some functions of the background dilaton,
which can be fine tuned to make the calculations more simple.
For instance, if one chooses these functions as
\beq
\alpha = - \frac{1}{2}\;\left( \phi + \ga  \right)
,\;\;\;\;\;\;\;\;\;\;\;\;
\beta=\frac{2-n}{n}
,\;\;\;\;\;\;\;\;\;\;\;\;
\rho = - \frac{1}{\phi + \ga}                          \label{2.6}
\eeq
then the bilinear part of the action $S+S_{gf}$ and the operator 
$\hat{H}$ have especially simple (minimal) structure
$$
\left(S + S_{gf}\right)^{(2)}
=\int d^4 x \sqrt{-g}\; \left( {\bar h}_{\mu\nu},\; h, \; \varphi \right)
\;\left( \hat{H} \right)\;
\left( {\bar h}_{\al\be},\; h, \; \varphi \right)^T
$$
\beq
\hat{H}=\hat{K}\Box+\hat{L}_{\la}\nabla^{\la}+\hat{M}    \label{2.7}
\eeq
Here $T$ means transposition, and the matrices in (\ref{2.7}) have the
form (we do not write the projectors to the symmetric traceless states
for brevity, but they have to be restored when the 
matrices are contracted)
$$
\hat{K}=\left(
\begin{array}{ccc}
\frac{\phi + \ga}{4} \;\delta^{\mu\nu ,\alpha\beta}  &  0  &  0 \\
0   & - \frac{2 - n}{8n} \; \left( \phi + \ga \right)   &  -(1 / 4) \\
0 & - (1 / 4) & \frac{1}{2(\phi +\ga)}-\frac{n-1}{\phi (n-2)}
\end{array}
\right)
$$
\vskip 1mm
\[
\hat{L}^{\lambda}=\left(
\begin{array}{ccc}
   \frac{1}{4} \delta^{\mu\nu, \alpha\beta} g^{\la\tau} +
\frac12 g^{\nu \beta}
\left( g^{\mu\tau} g^{\al\la} - g^{\al\tau} g^{\mu\la} \right)
   & - \frac{1}{4}\; g^{\mu\tau} g^{\nu\la}
   & - \frac{n-1}{\phi(n-2)}\; g^{\mu\tau} g^{\nu\la}  \\
                                                       \\
   \frac{1}{4}\; g^{\al\tau} g^{\be\la}
   & -\frac{n-2}{8n} \; g^{\la\tau}
   & \frac{n-1}{2n\phi}\;  g^{\la\tau}         \\
                                                 \\
   \frac{n-1}{(n-2)\phi} \;g^{\al\tau} g^{\beta \lambda}
   & \frac{1-n}{2n\phi}\; g^{\al\tau}
   & \left[\frac{n-1}{\phi^2(n-2)}-\frac{1}{2(\phi+\ga)^2}\right]
     g^{\la\tau}
\end{array}
\right) \phi_{\tau}
\]
\\
{\small
\[
\hat{M}=\left(
\begin{array}{ccc}
              \begin{array}{l}
\frac{1}{2} \left(\phi+\ga \right)
\left( R^{\mu\al\nu\be} - R^{\mu\al}g^{\nu\be} \right)+
\\
(1/2)\delta^{\mu\nu,\al\be} \left( {\Box}\phi \right)
- g^{\nu \beta}\left( \nabla^{\mu} \nabla^\alpha \phi  \right)
\\
- (1/4)(A+V)\;\delta^{\mu\nu,\al\be} + g^{\nu\be}A^{\mu\al}
           \end{array}
&
              \begin{array}{l}
\frac{n-4}{4n}\left(\na^\mu\na^\nu \phi - A^{\mu\nu}\right)
           \end{array}
&  \begin{array}{l}
\frac{n-1}{2(n-2)\phi^2}\;\phi^\mu\phi^\nu - \frac12 R^{\mu\nu}
              \end{array}
\\
\\
              \begin{array}{l}
\frac{n-2}{2n}\left(\na^\al\na^\be \phi \right) +
\frac{4-n}{4n}\;A^{\be\al}
           \end{array}
&
              \begin{array}{l}
\;\;\frac{(n-1)(4-n)}{4n^2}\left(\Box \phi \right)
\\
+ \frac{n-2}{8n} \left( \frac{n-4}{n}A+V \right)
           \end{array}
&
             \begin{array}{l}
- \frac{n-1}{4n}\;\frac{1}{\phi^2}\;\left(\na\phi\right)^2
\\
+ \frac{n-2}{4n} \;R + \frac{1}{4}\;V_1
           \end{array}
\\
\\
 \begin{array}{l}
\frac{1-n}{2(n-2)\phi^2}\;\phi^\al\phi^\be -
\frac12 R^{\al\be}
\\
+ \;\;\;\frac{n-1}{(n-2)\phi}\;\left(\na^{\al}\na^{\be}\phi\right)
              \end{array}
&
             \begin{array}{l}
\frac{n-1}{4n\phi^2}\left[\left(\na\phi\right)^2  -
2\phi\left(\Box\phi\right)\right]
\\
\,\,\,+ \frac{n-2}{4n} \;R + \frac{1}{4}\;V_1
           \end{array}
&
             \begin{array}{l}
\frac{n-1}{(n-2)\phi^3}\left[ \phi\Box\phi -
\left(\na\phi\right)^2 \right]
\\
\;\;\;\;\;\; + \frac{1}{2}\;V_2
           \end{array}
\\
\end{array}
\right)
\]
}
\vskip 1mm
\beq
A^{\mu\nu} =
\frac{n-1}{n-2}\;\left(\na^\mu\phi\right)
\left(\na^\nu\phi\right) + \left(\phi + \ga\right)\;R^{\mu\nu},
\; \; \; \; \; \; \; \; \;
A = A^{\mu\nu}\;g_{\mu\nu},
\; \; \; \; \; \; \; \; \;
\phi^\mu = \na^\mu\phi                        \label{2.8}
\eeq
To apply the Schwinger-DeWitt method we rewrite $\Tr\ln\hat{H}$
in the following way.
\beq
\Tr \ln{\hat {H}} \, = \, \ln Det{\hat {K}}\, + \,
\Tr\ln\left(\hat{1}\Box + {\hat {K}}^{-1} {\hat {L}}^{\mu}\nabla_\mu
+\hat{K}^{-1}\hat{M} \right)                            \label{2.9}
\eeq
The first term gives simple contribution to
the $1$-loop effective action, because it is nothing but an ordinary
functional determinant of the $c$-number matrix \footnote{Here
we abandon all the additive constants, an evident integrations
over the space-time variables and related $\sqrt{-g}$ factor, and also
the surface terms in the effective action.}.
\beq
\det {\hat K} = \frac{n-1}{8n}\,\frac{\ga(\phi+\ga)}{\phi} \label{2.10}
\eeq
The same concerns the last term in (\ref{2.3}) which can be
evaluated as
\beq
\det Y^{\mu\nu} = \phi + \ga                           \label{2.11}
\eeq
The second term in (\ref{2.9}) is  $\;\ln Det \;$ of the operator of the
standard
minimal form and it can be, in principle, evaluated with some accuracy
in the framework of the Schwinger-DeWitt method.
The bilinear form of the ghost action also has the minimal form
\beq
\hat{H}_{ghost}=
{ \de_\mu}^\nu\Box - \frac{1}{\phi+\ga}\;\phi^\nu\na_{\mu}
- \frac{1}{\phi+\ga}\;(\na_{\mu}\na^{\nu}\phi) + {R_{\mu}}^{\nu}
\label{2.12}
\eeq
and it's contribution can be also, in principle, evaluated with the
use of the standard technique.

Here we perform the derivation of the effective action with an accuracy
to the first order in curvature and in the corresponding second order
in the derivatives of the scalar field. This approximation corresponds
to the first coefficient
of the Schwinger-Dewitt expansion, which has, for the operator
\beq
{\hat H}_{min}
= {\hat 1}\Box + {\hat E}^\la \na_\la + {\hat D} \label{2.13}
\eeq
the form (see \cite{DTGF,dew,bavi} for full details)  
$$
\ln {Det} \left(- {\hat H}_{min} \right) =
- \int_{\epsilon^2}^{\infty}\,\frac{ds}{i\,s}\;\Tr U(x,x';s)
$$
\beq
U(x,x';s)
= e^{s\,{\hat H}_{min}} = U_0(x,x';s)\,
\sum_{n=0}^{\infty}\;{A}_n\,(i\,s)^n                \label{SdW}
\eeq
The trace of the first $A_1(x,x)$ coefficient is defined, for the operator
${\hat H}_{min}$, as
\beq
\tr A^{min}_1 = \int d^4x \sqrt{-g}\,
\tr\, \left\{ {\hat D} - \frac14\;{\hat E}^\la {\hat E}_\la
+ \frac{{\bar 1}}{6}\;R \right\}                      \label{2.14}
\eeq
Using (\ref{2.3}), (\ref{2.9}), (\ref{2.13}) and (\ref{2.14}), after
some algebra, we arrive at the following expression
\beq
\Gamma^{(1-loop)}_{1} =
\mu^{n}\,\int d^n x\sqrt{-g}\;
\left\{\, {\cal A}(\phi)\;g^{\mu\nu}\;\partial_{\mu}\phi\; 
\partial_{\nu}\phi +
{\cal B}\,(\phi)R + {\cal C}(\phi) \, \right\}             \label{2.15}
\eeq
where $\mu$ is dimensional parameter related to $\epsilon$ in (\ref{SdW}), 
and
$$
{\cal A}(\phi) =
\frac{1}{8\,\left(n - 2 \right) \,
\left(n - 1 \right)\, \left( \ga + \phi \right)^2}\;
\left\{ \frac{\ga^2}{\phi^2} \left(6n - 4 - 2 n^2 \right) +
\frac{\ga}{\phi}  \left(16  - 28n + 4{n^2} + 12{n^3} - 4{n^4} \right) -
\right.
$$
$$
\left.
- 26\,n + 49\,{n^2} - 30\,{n^3} + 7\,{n^4} +
\frac{\phi}{\ga}  \; \left( 16 - 28\,n + 16\,{n^2} - 2\,{n^3} \right)
\right\}
$$
\vskip 1mm
$$
{\cal B}(\phi) = \frac{1}{12}\,\left( 3\,n - 5\,n^2 - 1 \right)
 + \frac{ n\,\phi}{2\,\ga\,( n - 1) }
\;\;\;\;\;\;\;\;\;\;\;\;\;\;\;\;\;\;\;\;\;\;\;\;\;\;\;\;\;
$$
\vskip 1mm
\beq
{\cal C}(\phi) =
{{{\la}\,{n}\,{\phi^{{2\over {n - 2}}}}}\over {\ga\,( n^2 - 3\,n + 2 ) }} 
\;\left(
\,n \, \phi + \phi + \ga   \right) +
 {{n\,\left( - \ga + \ga\,{n^2} + \phi \right) \,
      \left( {\la}\,{\phi^{{n\over {n - 2}}}} +
        {\ga^{{n\over {n - 2}}}}\,{\tau} \right) }
     \over {2\,\ga\,\left( 1 - n \right) \,
      \left( \ga + \phi \right) }} \;\;\;\;\;\;\;\;\;
\label{res}
\eeq
The above result (\ref{res}) is the lowest (in background dimension)
part of the one-loop correction to the classical action.
It is easy to see that (\ref{res}) doesn't satisfy the conditions
imposed by conformal duality, and therefore the quantum corrections
violate the symmetry in this approximation. At the same time,
as it will be shown below, the expression (\ref{res})
contains a big gauge fixing arbitrariness which can be fixed if only one
uses the equations of motion.

\vspace{5mm}

\noindent{\large \bf 4. Gauge fixing independent part of quantum 
corrections}
\vskip 1mm

The analysis of (\ref{res}) can be performed in a reasonable way only
after we explore it's gauge fixing dependence.
Such a dependence for the off-shell effective action takes place for any
gauge theory. Fortunately it can be investigated in general form
\cite{dew-67,vlt} (see also \cite{book}).
For the one-loop contribution to the effective action one can easily find
this dependence explicitly, following the
method of \cite{bavi} (see also it's development in \cite{avr,shja}
for the gauge fixing condition with an additional weight operator 
(\ref{2.4})).
From the general expression
(which is derived, for instance, in an Appendix of \cite{shja})
it follows that
if the weight operator $Y^{\mu\nu}$ and the gauge fixing condition
$\chi_{\mu}$ depend on arbitrary\\parameter $t$, then the derivative
of (\ref{2.3}) on $t$ is
\beq
\frac{d}{dt}
\Gamma^{(1-loop)}_{1} =
 -{1\over2}\;G^{nk}\;\varepsilon_p\;
{ {\delta}{\nabla}^p_{\alpha} \over{\delta w^n}}\; M^{\alpha\mu}
\left[\, { {\delta\chi^{\nu}} \over {\delta w^k} }\;{Y'_{\mu\nu}} +
2\,Y_{\mu\nu}\;
\left( { {\delta\chi^{\nu}} \over {\delta w^k} } \right)'\,\right]  
\label{2.16}
\eeq
where the touch denotes the derivative on $t$, $\;M^{\alpha\mu}$ and
$G^{nk}$ are the propagators of the gauge ghosts and
fields $w^p = \left( {\bar h}_{\mu\nu}, h, \varphi \right)$
correspondingly, $\na^p_{\al}$ are the generators of the field $w^p$ and
$\varepsilon_p = \de S\,/ \,\de w^p$ are classical equations of motion
(B.1), (B.2).
Taking $t$ to be, in turn, the gauge fixing parameters 
$\al,\be,\rho$  and
integrating over these parameters one arrives at the complete explicit
form of the gauge fixing dependence of the one-loop effective action.

Thus at one loop the  dependence of $\Gamma^{(1-loop)}_{1}$
on the gauge parameters $\al,\be,\rho$ is proportional to the
equations of motion (B.1), (B.2). These equations have the same dimension
as the integrand of $\Gamma^{(1-loop)}_{1}$, and therefore the gauge fixing
dependence is rather strong in this case. In particular all the functions
${\cal A}(\phi), {\cal B}(\phi), {\cal C} (\phi)$ can be essentially
changed for different values of
$\al, \be, \rho$, and (as will be shown below) only one combination of
these functions is gauge independent.
One can supposes that this rate can be changed if we take
into account the higher orders in the Schwinger-Dewitt expansion, where
the integrands of $\;\Tr a_{k}(x,x)\,$ have higher dimension.

In order to extract some 
invariant quantity from the expression (\ref{2.15})
one can assume that the background fields $g_{\mu\nu}$ and $\phi$
satisfy classical equations of motion (B.1), (B.2). If fact, due to the
dimensional reasons, we need only the trace of the equation for the
metric, so instead of (B.1), (B.2) one can use (B.3), (B.4).
One can directly substitute (B.3) into (\ref{2.15}), however in order
to use (B.4) one needs first to perform some transformations. From (B.4)
it follows that for any function $F(\phi)$ and for any solution of the
equations of motion (B.3), (B.4)
\beq
\int d^nx \sqrt{-g}\; F(\phi)\;\left\{ \frac{1}{\phi}
\left(\na \phi \right)^2 - 2\left(\Box \phi \right) - \phi S(\phi)
\right\} = 0                                \label{2.17}
\eeq
Integrating by parts we arrive at
\beq
\int d^nx \sqrt{-g}\;
\left\{ \frac{1}{\phi} F(\phi) + 2 F_1(\phi) \right\}
\left(\na \phi \right)^2
= \int d^nx \sqrt{-g}\; F(\phi)\;\phi S(\phi)  \label{2.18}
\eeq
The {\it lhs} of (\ref{2.18}) is invariant under the change
$F(\phi) \rightarrow F(\phi) + C_1 \phi^{-1/2}$ with $C_1 = const$.
In fact this means that for any solution of (B.3), (B.4) the
function $C_1 \phi^{-1/2}$ possesses the property
\beq
\int d^nx \sqrt{-g}\; C_1 \phi^{-1/2}\;\phi S(\phi) = 0  \label{2.19}
\eeq
and therefore can be always safely omitted. Now we remind (\ref{2.15})
and put
\beq
\frac{1}{\phi} F(\phi) + 2 F_1(\phi) = {\cal A}(\phi)
\label{2.20}
\eeq
The solution of the last equation has the form
\beq
F(\phi) =  C_1 \phi^{-1/2} + \phi^{-1/2}\,\int_{\phi_0}^{\phi}
d\phi\;\phi^{1/2}\; {\cal A}(\phi)
\label{2.21}
\eeq
When (\ref{2.21}) is substituted into (\ref{2.18}) and then to
(\ref{2.15}), one has to take into account (\ref{2.19}), that makes
the values of constants $C_1,\,\phi_0$ irrelevant. Finally we arrive at
the following expression for the gauge-fixing independent on-shell
effective action:
$$
\Gamma^{(1-loop)}_{1,\,ms} =
\mu^{n}\,\int d^nx\sqrt{-g}\;
\left\{
\frac{n\,\tau}{2\,(n-2)}\,\gamma^{2 \over {n-2}} \,
\left( \frac{5n^2 - 3n + 22}{6} - \frac{n}{(n-1)}\,\frac{\phi}{\ga} 
\right) + 
\right.
$$$$
\left.
+ V_1(\phi)\,\left( \frac{\phi}{\ga} -
\frac{1}{n-1} \right)
- \frac{n\,V(\phi)}{2\,(n-1)\,(\phi + \ga)}\;
\left( n^2 - 1 + \frac{\phi}{\ga} \right) +
\right.
$$$$
\left.
+ \frac{n\,S(\phi)}{2\,(n-1)}\;\left[
\,1 - \frac{2\,(n^3 - 2 n^2 - 4 n + 6)}{4\,(n-2)}\;\frac{\phi}{\ga} +
+ \frac{2\,n^3 - 3\,n^2 - 2\,n + 2}{(n-1)}\;
\left( \frac{\phi}{\ga} -
\sqrt{ \frac{\phi}{\gamma} } \,
\arctan \sqrt{ \frac{\phi}{\gamma}} \;\right)
\right.  \right.
$$
\beq
\left.\left.
+ \frac{11\,n^4-40\,n^3+27\,n^2+36\,n-36}{4\,(n-1)\,(n-2)}\;
\left(- \frac{\phi}{\phi + \ga} +
\sqrt{ \frac{\phi}{\gamma} } \,
\arctan \sqrt{ \frac{\phi}{\gamma}} \;\right)
\right]\,\right\}                                      \label{mass}
\eeq
The $\arctan$'s appear because of integration in (\ref{2.21}). One can
see that all the divergences depend on
the potential through
$\frac{V(\phi)}{\phi + \ga}$, on it's first
derivative $V_1(\phi)$, on function $S(\phi)$, which is defined in (B.4)
and also on the ratio $(\phi / \ga)$. The on-shell expression 
possesses the
homogeneity under the simultaneous rescaling of scalar and constant $\ga$
\beq
\phi \rightarrow k\,\phi,\;\;\;\;\;\;\;\;\;\;\;\;\;\;\;\;\;\;\;\;
\ga \rightarrow k\,\ga,\;\;\;\;\;\;\;\;\;\;\;\;\;\;\;\;\;\;\;\;k = const
\label{2.22}
\eeq
just as the corresponding on-shell expression for the classical action
\beq
S_{ms} = \frac{1}{\ka^2}\,\int d^nx\,\sqrt{-g}\;
\left\{
\frac{\tau}{n-2}\,\gamma^{2 \over {n-2}} \,(n\,\phi - 2\,\ga)
- \frac{\la\,(n+2)}{n-2}\,\phi^{n \over {n-2}} \right\}  \label{2.23}
\eeq
One can see that the order of homogeneity for (\ref{mass}) is different
from the one for (\ref{2.23}) (the exception is only singular case
$n=2$, see Appendix C) and therefore the on-shell part of
$\Ga^{(1-loop)}_1$ is not invariant under the transformation of conformal
duality (\ref{n14}). 
In fact the violation of the conformal duality in this
approximation is not a wonder because the similar thing happens for
conformal scalar field. The $a_1$ coefficient can not be conformal
invariant because of it's dimension with the only one exception of 
$n=2$.

\vspace{5mm}

\noindent{\large \bf 5 Conclusion}
\vskip 1mm

We have formulated the conformal transformation and conformal
duality in the metric-dilaton models in arbitrary dimension,
and derived the lower-order one-loop correction to the effective
action for the theory with conformal duality. For odd dimensions 
the one-loop effective
action is free of logarithmic divergences and one can
consider the above result as an approximation to the effective
action in the weak coupling and weak gravitational field limit.
The conformal duality can link the strong and weak
gravitational field regimes for the metric-dilaton theories.
The first approximation which we have explored here,
shows the general structure of quantum corrections, but unfortunately 
it provides too little information because of the strong dependence
on the gauge-fixing arbitrariness. 
The only one part of the one-loop correction to the classical action,
is the expression (\ref{mass}) which results from the substitution 
of the classical equations of motion into the first coefficient 
of the Schwinger-DeWitt expansion. In the regime of weak 
gravitational field
the parameters of such an expansion are i) the dimensional coupling 
constant $\ka$ and ii) the curvature tensor of the background metric.
If we regard both quantities as small parameters, then the quantum 
correction is small too. In this case one can perform the finite 
renormalization of the metric and scalar field in such a way that,
in terms of the new fields, all but the last (potential) terms in the
action have the "classical" form (\ref{2.1}). If
we accept this procedure then the only quantum contribution will be
the correction to the potential $V(\phi)$ in (\ref{2.1}). 
This correction is indeed defined by the expression (\ref{mass}) with
additional factor of $\ka$, therefore this correction is gauge
fixing and parametrization independent. 
Thus derived, the effective potential can be used to extract the
information about the strong curvature regim\footnote{The quantum 
correction leads to the soft breaking of dual symmetry. The source 
of this breaking is the the loop expansion of the effective action. 
Consequently if we take into 
account quantum correction, the procedure of going to dual regim may
lead to the nontrivial change of the potential.}, hence it
enables one to explore how 
quantum corrections change the expansion rate of the
early Universe, or how they affect the black hole solutions in the
vicinity of singularity. We hope to investigate these problems in a
close future.

Indeed it should be interesting also to use
a more complicated methods of calculations (see for example 
\cite{avr-exp}, \cite{zeta}) to the 
theories with conformal duality. These methods can help to extract 
the nonlocal part of the effective action, which can be free of
the gauge arbitrariness, and can be directly 
applicable to cosmology and black 
hole physics. On the other hand, one can explore the functional 
properties of
the operators resulting from the conformally dual theory, as it has been
done for the conformal symmetry itself (see, for example, \cite{bran}).

\vspace{5mm}

\noindent{\large \bf Acknowledgments}
\vskip 1mm

Author is grateful to I.G. Avramidi and M.S. Plyushchay
for helpful discussions and also thanks for the warm
hospitality the Departamento de Fisica Teorica en Universidad
de Zaragoza and the Departamento de Fisica en Universidade Federal de
Juiz de Fora, where this work has been started and completed,
and also CICYT for financial support
under grant AEN94-0218 and MEC-DGICYT for fellowship.
The work was supported in part by Russian Foundation for Basic
Research under the project No.96-02-16017.

\vskip 15mm

\noindent{\large \bf Appendix A. Soft breaking of conformal symmetry in
$n$ dimensions.}
\vskip 1mm

Following \cite{conf} one can consider the possibility of the soft
breaking of the conformal symmetry of the action $S_{B(\phi),\la}$
(\ref{n6}) in $n$ dimensions.
For this purpose we need the conformal Noether identity, which has
exactly the same form as in $n=4$
$$
B_1(\phi)\; g_{\mu\nu} \;\frac{\delta S}{\delta g_{\mu\nu}} -
B(\phi)\;\frac{\delta S}{\delta \phi} = 0               \eqno{(A.1)}
$$
where the factor $-\frac{2B}{\phi B_1}$ stands for the
 conformal weight of the scalar field $\phi$  which depends on the form
of the function $B(\phi)$.
It is an analog (one can say generalization) of the
conventional conformal weight $\frac{2-n}{2}$ of the scalar field
in (\ref{n8}). The eq. (A.1) is the operator form
of the symmetry transformation (\ref{n4}).
According to (A.1) the equations of motion for the scalar field and
for the conformal factor
of the metric are linearly dependent. For the special case of GR one
meets $B=const$ and both therms in (A.1) vanish. Consider the
nonconstant  $B$.

The soft breaking of the conformal symmetry means that the functions
$A$ and $B$ satisfy the symmetry condition (\ref{n6})
whereas the restrictions on the potential term $C$
are not imposed. In such a case, however, an arbitrary function
$C(\phi)$ satisfies the differential equation, which results
from the equations of motion \cite{conf}
$$
C_1(\phi)B(\phi) = \frac{n}{n-2}\;C(\phi)B_1(\phi)   \eqno(A.2)
$$
that has the unique nonzero
solution $C(\phi) = constant \cdot \left[B(\phi)\right]^{\frac{n}{n-2}}$.
One can easily check that this statement
is correct even if we add the action of matter, if this matter does not
depend on the field $\phi$.
Thus, just as in $n=4$ case \cite{conf}, only the symmetric
form of $C(\phi)$ is consistent with the equations of motion.
It is interesting to note that the soft breaking of conformal
symmetry is possible (at least in the sense we are considering it
here) for the both types of the higher derivative conformal
metric-dilaton models \cite{shja} and \cite{ait,conf}.

\vspace{5mm}

\noindent{\large 
\bf Appendix B. Equations of motion for the dual theory.}
\vskip 1mm

Let us consider the equations of motion for the theory with conformal
duality. For the sake of simplicity we start with the special simple
case (\ref{2.1}) which is related with the general one (\ref{n11})
by i)conformal transformation of the metric ii) reparametrization of the
scalar field. The dynamical equations for the theory (\ref{2.1}) have
the form
$$
\frac{\de S}{\de g^{\mu\nu}} = \frac{n-1}{n-2}\;\frac{1}{\phi}\;
\left( \frac12\;(\na \phi)^2\,g_{\mu\nu}
 - \partial_\mu\phi\;\partial_\nu\phi \right)
+ (\na_\mu\na_\nu\phi) - ({\Box}\phi)g_{\mu\nu} -
$$
$$
-(\phi + \ga) \left(\frac12 \,R\,g_{\mu\nu}
- R_{\mu\nu} \right) + \frac12\;V\,g_{\mu\nu} = 0
\eqno(B.1)
$$
\vskip 1mm
$$
\frac{\de S}{\de \phi} = \frac{n-1}{n-2}\;\left[
\;\frac{(\na\phi)^2}{\phi^2} - \frac{2}{\phi^2}\;(\Box \phi)
\right] + \la \;\frac{n}{n-1}\;\phi^{\frac{2}{n-2}} + R = 0
\eqno(B.2)
$$
Taking trace of (B.1) and comparing it with (B.2) we find the constraint
$$
R = \frac{n\,\tau}{2-n}\; \ga^{\frac{2}{n-2}}
\eqno(B.3)
$$
Substituting (B.3) back to (B.2) we obtain
$$
\frac{1}{\phi^2}\;(\na\phi)^2 - \frac{2}{\phi^2}\;(\Box \phi)
= \frac{n}{n-1}\;\left( \tau\,\ga^{\frac{2}{n-2}} -
\la\,\phi^{\frac{2}{n-2}} \right) = S(\phi)
\eqno(B.4)
$$
Since the equation for $\phi$ is factorized it is not very difficult
to solve it for some special cases. For instance, if we are looking for
the spatially homogeneous solutions and $\phi$ depends
only on time, (B.4) becomes
$$
\left( \phi' \right)^2 - 2 \phi \phi'' =
\frac{n}{n-1} \; \phi^2 \left( \tau\, \ga^{\frac{2}{n-2}} -
\la\, \phi^{\frac{2}{n-2}} \right)
\eqno(B.5)
$$
that can be easily solved. Along with the obvious constant solution
$$
\phi = \ga \left( \frac{\tau}{\la} \right)^{\frac{n-2}{2}}
\eqno(B.6)
$$
there is a nonconstant one which can be presented in the integral form
$$
t - t_0 = \pm
\int_{\phi_0}^\phi
\frac{d\phi}{ \left\{ C\phi + \phi^2 \left[
\frac{n\,\tau}{n-1} \, \ga^{\frac{2}{n-2}} - \frac{n-2}{n-1} \,
\la \phi^{\frac{2}{n-2}} \right] \right\}^{1/2} }            \eqno(B.7)
$$
where $C$ is integration constant, which should be, in principle,
defined on physical backgrounds.
The last integral is elementary for the special case $n=3,4;\;\;C=0$.
In particular, for $n=3$ we find
$$
\phi = \ga \left( \frac{\la}{3\tau} \right)^{1/2}
\left[ 1 + \tan^2 \left(
A + |\ga|\sqrt{3\tau}(t-t_0)\right) \right],\;\;\;\;\;\;\;\; A = const
\eqno(B.8)
$$
In more general case $C\neq 0$ eq. (B.7) can be evaluated within some
approximation scheme.
For example, if we are interested in the cosmological solution for which
the dilaton becomes constant in the limit $t \rightarrow + \infty$
then the fixed points method can be applied, and we need to find the
values of dilaton which illuminate the denominator of the integrand
in (B.6). It is useful to rewrite the condition
$$
C\phi +
\phi^2 \left[
\frac{n\,\tau}{n-1}\,\ga^{\frac{2}{n-1}} - \frac{n-2}{n-1}\,
\la\phi^{\frac{2}{n-2}}\right] = 0                 \eqno(B.9)
$$
in the algebraic form via the new variable $\psi$, where
$\phi = \psi^{n-2}$. The resulting equation
$$
\psi^{n-2}\left[ C +
\frac{n\,\tau}{n-1}\,\ga^{\frac{2}{n-2}} - \frac{n-2}{n-1}\,
\la\phi^{n-2} \right] = 0                          \eqno(B.10)
$$
can have a nonzero solutions. It is remarkable that the
fixed point $\psi = \phi = 0$ is stable at $t \rightarrow + \infty$
only for the even space-time dimensions $n$.

Of course the simple constraint (B.3) and equation (B.4) hold only
for the special model (\ref{2.1}) while for the general theory with
conformal duality (\ref{n11})
scalar curvature is not constant and the equation for
scalar is not so simple. However one can always obtain
the corresponding quantities in the general (\ref{n11}) with the use
of conformal transformation and reparametrization of the special model
(\ref{2.1}), that is using a version of solution-generating technique
\cite{bek}.

\vspace{5mm}

\noindent{\large \bf Appendix C. Special case of two-dimensional theory.}
\vskip 1mm

The actions for both conformal metric-dilaton gravity (\ref{n6})
and for the theory with conformal duality (\ref{n12}), (\ref{n13})
become singular at $n \rightarrow 2$, and therefore two dimension is
special case, 
where those symmetries can not be realized in their literal
form. Of course we know that the conformal symmetry can be realized
in $n = 2$, and therefore it is interesting to see how the correspondence
can be achieved. This may be useful, for instance, in the framework of
$2 + \varepsilon$ approach to quantum gravity (see \cite{AKKN} for
discussion and references).
First we start from (\ref{2.1}) and change the scalar
variable $\phi$ to another one $\chi$ as
$$
\phi = \frac{n-2}{2\,(n-1)}\;\chi^2               \eqno(C.1)
$$
In terms of new variable the action becomes
$$
S = \int d^nx \sqrt{-g}\; \left\{ \frac12\,g^{\mu\nu}\,\partial_{\mu}\chi
\partial_{\nu}\chi +
\frac{n-2}{4\,(n-1)}\,\chi^2\,R + \ga\,R +
\la\;\left(\frac{n-2}{n-1}\right)^{ \frac{n}{n-2} }
\,\chi^{\frac{2\,n}{n-2}}
+ \tau \ga^{\frac{2\,n}{n-2}}     \right\}          \eqno(C.2)
$$
The conformal model can be extracted from (C.2) $\,\,$ (just as from
 (\ref{2.1})) when one puts $\tau = \ga = 0$.

In the limit $n \rightarrow 2$ we meet the following behaviour of the model.
The only remaining dynamical term is $g^{\mu\nu}\,\partial_{\mu}\chi
\partial_{\nu}\chi$ , the $\ga\,R$
term becomes topological 
and $\chi^2\,R$ one disappears. One can easily check
that the potential term tends to zero, while the cosmological constant
$\tau \ga^{\frac{2\,n}{n-2}}$ becomes infinite. No any dual symmetry is
observed in this limit, while the conformal symmetry is restored
at $\tau = \ga = 0$.

One can perform another singular 
(at $n=2$) change of variables in the action
(\ref{2.1}). Let's put
$$
\phi^\frac{n}{n-2} = \psi                    \eqno(C.3)
$$
Then the action (\ref{2.1}) becomes
$$
S = \int d^nx \sqrt{-g}\; \left\{ \frac{(n-1)\,(n-2)}{n^2}\,
\psi^{-\frac{n+4}{n}}
g^{\mu\nu}\,\partial_{\mu}\psi \partial_{\nu}\psi +
\psi^\frac{n-2}{n}\,R + \ga\,R +
\la\;\psi + \tau \ga^{\frac{2\,n}{n-2}}     \right\}          \eqno(C.4)
$$
In these variables, just as in (C.2), the limit $n \rightarrow 2$ is
singular
in the cosmological term only. However, contrary to (C.2), the dynamical
scalar term disappears. Therefore the dynamics of the theory
(\ref{2.1}) at $n \rightarrow 2$, which we have met in (C.2), is caused
by the
singular change of variables (C.1), and such a dynamics doesn't take
place for another choice of the variables.
Thus we see that one can perform such a limit in a different ways and
arrive at
the essentially different $2d$ metric-dilaton models. The reason for
this is that  the original
model  (\ref{2.1}) is singular at $n \rightarrow 2$.

\vspace{5mm}

\noindent{\large \bf Appendix D.
The elements of the background field method.}
\vskip 1mm

Here we collect
some intermediate formulas related with the bilinear expansion
of the action (\ref{2.1}) in the background field method.
One can find an expansion
of scalar curvature and $\sqrt{- g}$,
corresponding to (\ref{2.2}), in many papers or in \cite{book}, 
so we do not write them here. For the kinetic term we use
$$
\frac{1}{\phi'} = \frac{1}{\phi\, (1 + \varphi / \phi)} =
\frac{1}{\phi} - \frac{\varphi}{\phi^2} + \frac{\varphi^2}{\phi^3} +
...                                                  \eqno(D.1)
$$
and after some integrations by parts and a little tedious algebra 
arrive at
$$
S^{(2)} =
{S^{(2)}}_{(\varphi,\varphi)} +
{S^{(2)}}_{(\varphi,{{\bar h}_{\al\be}})} +
{S^{(2)}}_{({{\bar h}_{\mu\nu}},\varphi)} +
{S^{(2)}}_{(\varphi,h)} +
$$$$
{S^{(2)}}_{(h,\varphi)} +
{S^{(2)}}_{({{\bar h}_{\mu\nu}},{{\bar h}_{\al\be}})}+
{S^{(2)}}_{(h,{{\bar h}_{\al\be}})}+
{S^{(2)}}_{({{\bar h}_{\mu\nu}},h)}+
{S^{(2)}}_{(h,h)}                              \eqno(D.2)
$$
where
$$
{S^{(2)}}_{(\varphi,\varphi)} =
\int d^nx \,\sqrt{-g} \;
\varphi \left[ - \frac{a}{\phi}\; {\Box}
+ \frac{a}{\phi^2}\; (\na^{\mu}\phi)\na_\mu
- \frac{a}{\phi^3}\;(\na \phi)^2
+ \frac{a}{\phi^2}\;({\Box}\phi) + \frac{1}{2}\;V_2
\right] \varphi
$$$$
{S^{(2)}}_{(\varphi,{{\bar h}^{\al\be}})}
= \int d^nx \,\sqrt{-g} \;
\varphi \left[ \frac12\;\na_\al\na_\be
+ \frac{a}{\phi}\; \phi_\al\na_\be
- \frac{a}{2\,\phi^2}\;\phi_\al\phi_\be
+ \frac{a}{\phi}\;(\na_\al\na\be \phi)
- \frac12\,R_{\al\be}
\right]
{\bar h}^{\al\be}
$$$$
{S^{(2)}}_{({{\bar h}_{\mu\nu}},\varphi)}   = \int d^nx \,\sqrt{-g} \;
{\bar h}^{\mu\nu}
   \left[ \frac12\;\na_\mu\na_\nu
- \frac{a}{\phi}\; \phi_\mu\na_\nu
+ \frac{a}{2\,\phi^2}\;\phi_\mu\phi_\nu
- \frac12\,R_{\mu\nu}
\right]\varphi
$$$$
{S^{(2)}}_{(\varphi,h)}  = \int d^nx \,\sqrt{-g} \;
 \varphi  \left[
\frac{1-n}{2\,n}\;{\Box}
+ \frac{n-1}{2\,n\,\phi}\; \left( - \phi_\la\na_\la
 + \frac{1}{2\;\phi}\; (\na\phi)^2
  - ({\Box}\phi)
+ \frac{\phi}{2\,a}\;R \right)
+ \frac{1}{4}\;V_1
\right] h
$$$$
{S^{(2)}}_{(h,\varphi)}  = \int d^nx \,\sqrt{-g} \;
h \left[\frac{1-n}{2\,n}\;{\Box}
+ \frac{n-2}{2\,n}\; \left( \frac{a}{\phi}\; \phi_\la\na_\la
 - \frac{1}{2}\; \frac{a}{\phi^2}\; (\na_\phi)^2
 + \frac{1}{2}\;R \right)
+ \frac{1}{4}\;V_1
\right] \varphi
$$$$
{S^{(2)}}_{({{\bar h}_{\mu\nu}},{{\bar h}_{\al\be}})}  =
\int d^nx \,\sqrt{-g} \;  {{\bar h}_{\mu\nu}}
   \left[ \frac14\,\de_{\mu\nu,\al\be}\,\left((\phi + \ga)\,{\Box} +
\phi^\la\na_\la + 2\,({\Box} \phi) - A - V \right) +
\right.
$$$$
\left.
+ \frac12\,g_{\nu\be}\,\left( - \na_\al\na_\mu - \phi_\al\na_\mu -
2\,(\na_\al\na_\mu \phi) + A_{\al\mu}  \right)
\right] {{\bar h}_{\al\be}}
$$$$
{S^{(2)}}_{(h,{{\bar h}_{\al\be}})}  = \int d^nx \,\sqrt{-g} \;
 h \left[\frac{n-2}{4\,n}(\phi + \ga)\na_\al\na_\be +
\frac{n-1}{2\,n}\phi_\al\na_\be +
\frac{n-2}{2\,n}(\na_\al\na_\be \phi) +
\frac{4-n}{4\,n}\,A_{\al\be}
\right] {{\bar h}_{\al\be}}
$$$$
{S^{(2)}}_{({{\bar h}_{\mu\nu}},h)}  = \int d^nx \,\sqrt{-g} \;
 {{\bar h}_{\mu\nu}}  \left[
\frac{n-2}{4\,n}\,(\phi + \ga)\,\na_\mu\na_\nu -
\frac{1}{2\,n}\,\phi_\mu\na_\nu +
\frac{n-4}{4\,n}\;(\na_\mu\na\nu\phi) +
\frac{4-n}{4\,n}\,A_{\mu\nu}
\right] h
$$$$
{S^{(2)}}_{(h,h)}  = \int d^nx \,\sqrt{-g} \;
 h  \;\left(\frac{n-1}{4\,n^2} \right)
\,\left[ (2-n)\,(\phi + \ga) {\Box} +
(2-n)\; \phi_\la\na_\la + (4-n)\; ( {\Box} \phi)
\right] h                                          \eqno(D.3)
$$
where $a = \frac{n-1}{n-2}$.
Supplemented by the gauge fixing term (\ref{2.4}) -- (\ref{2.6})
the expressions (D.2), (D.3) lead to the minimal matrix form of
$S^{(2)} + S_{gf}$ in eq. (\ref{2.7}).

\newpage

\end{document}